\documentclass[english,aip,jmp, eqsecnum]{revtex4-1}
\usepackage{mathptmx}
\usepackage[T1]{fontenc}
\usepackage[latin1]{inputenc}
\usepackage{bm}
\usepackage{esint}

\makeatletter

\providecommand{\tabularnewline}{\\}

 \@ifundefined{textcolor}{}
 {%
   \definecolor{BLACK}{gray}{0}
   \definecolor{WHITE}{gray}{1}
   \definecolor{RED}{rgb}{1,0,0}
   \definecolor{GREEN}{rgb}{0,1,0}
   \definecolor{BLUE}{rgb}{0,0,1}
   \definecolor{CYAN}{cmyk}{1,0,0,0}
   \definecolor{MAGENTA}{cmyk}{0,1,0,0}
   \definecolor{YELLOW}{cmyk}{0,0,1,0}
 }

\makeatother

\usepackage{babel}
\begin{document}

\title{Resolution of unity for fermionic Gaussian operators }

\author{Laura E. C. Rosales-Z\'arate}
\affiliation{Centre for Atom Optics and Ultrafast Spectroscopy, Swinburne University
of Technology, Melbourne 3122, Australia}

\author{ P. D. Drummond}
\affiliation{Centre for Atom Optics and Ultrafast Spectroscopy, Swinburne University
of Technology, Melbourne 3122, Australia}
\address{pdrummond@swin.edu.au}

\begin{abstract}
The fermionic Gaussian operator basis provides a representation for
treating strongly correlated fermion systems, as well as playing an
important role in random matrix theory. We prove that a resolution
of unity exists for any even distribution of eigenvalues over hermitian
fermionic Gaussian operators in the nonstandard symmetry classes.
This has some important consequences. It demonstrates a useful technique
for constructing fundamental mathematical identities in an exponentially
complex Hilbert space. It also shows that, to obtain nontrivial results
for random matrix canonical ensembles in the nonstandard symmetry
classes, it is necessary to consider ensembles that are not even functions
of the eigenvalues. We show that the same restriction does not apply
to the standard Wigner-Dyson symmetry classes of random matrices.
\end{abstract}
\maketitle

\section{Introduction}

The linear transformations of fermion operators that preserve anti-commutation
relations is an important fundamental symmetry group. It is known
to have a representation as an exponential of a general quadratic
form in the fermion operators. By analogy with the Gaussian function
in statistics, we term this a fermionic Gaussian operator. The group
properties and integration measures of these operators have been extensively
investigated in the mathematical physics literature\cite{Balian_Brezin_Transformations,Altland_Zirnbauer:1997,Dyson:1962_Threefold,Dyson:1962_JMP140,Dyson:1970_CommMathPhys,Weyl_book_groups}.
They have been used to generate the generalized fermionic coherent
states\cite{Perelomov_book_Coherent_state,Review_coherent_states_Gilmore_1990,Gilmore_tutorial_CS_bosons_fermions_1983}.
This group is also known to correspond to a general symmetry class
in random matrix theory\cite{Altland_Zirnbauer:1997,Forrester_book:2010,Caselle_Magnea:2004_ReviewRMT_SymSpaces}.

The Gaussian fermion operators have other fundamental applications.
They are an operator basis for all fermionic density operators. Both
the well-known thermal states and the BCS states are examples of this.
The Gaussian fermionic phase-space representation\cite{Corney_PD_JPA_2006_GR_fermions,Corney_PD_PRB_2006_GPSR_fermions,Corney_PD_PRL2004_GQMC_ferm_bos}
makes use of this property to generate a positive distribution over
such operators. This is a complete phase-space representation for
either fermionic and bosonic many body density operators. It has been
used, for example, to computationally evaluate the ground state of
the fermionic Hubbard model \cite{Imada_2007_GBMC}and to obtain an
expression for the linear entropy of a density operator \cite{PRA_Entropy_paper_2011}. 

In this communication, we show that an expansion using hermitian fermionic
Gaussian operators provides a continuous positive-definite resolution
of the fermionic identity operator. Our result holds for any integrable
even distribution of eigenvalues. This resolution is complete, in
that it generates an identity operator for the entire Hilbert space,
not a subset. The proof of the resolution of unity is given in terms
of the polar coordinates of skew symmetric matrices. This shows that
the fermionic Gaussian operators can be thought of as coherent operators.
In terms of continuity and completeness, they have an analogous role
to the bosonic coherent state projectors \cite{Glauber_1963_P-Rep,Klauder_book_CS}. 

To understand this comparison, we recall that one of the most useful
properties of the bosonic coherent states is their resolution of unity\cite{Klauder_1960_AnnalsPhys,Glauber_1963_P-Rep,Sudarshan_1963_P-Rep}.
This resolution of the bosonic identity operator uses positive-definite
coherent state projection operators, with a positive measure over
complex amplitudes. It is the basis for such well-known methods as
the Husimi Q-function \cite{Husimi1940}. Our resolution of unity
for the fermionic Gaussian operators has similar properties, with
a normalized Gaussian operator replacing the coherent-state projector
in the expansion of the unity. Just as in the boson case, a resolution
of unity is the foundation for calculating other fermionic identities
and phase-space representations. These will be treated in subsequent
publications. 

An alternative approach by Gilmore \cite{Gilmore:1972,Gilmore_RMF1974_Properties_CS}
and Perelomov \cite{Perelomov_1972_Coherent_states_LieG,Perelomov_GCS_bosons}
defined coherent state representations for fermions using generalized
fermionic coherent states. These are obtained as the action of a representation
of a Lie group on an extremal state. This definition is related to
the dynamical structure of the corresponding creation and annihilation
operators, using $U(N)$ Lie group methods \cite{Perelomov_book_Coherent_state,Gilmore_tutorial_CS_bosons_fermions_1983,Review_coherent_states_Gilmore_1990,Klauder_book_CS,Blaizot_PRC_1981,Blaizot_Book,Rowe_Ryman_coherent_states,Rowe1991_PRC44_PairCouplingTheo_FS,Coherent_states_book}.
In this approach, an integral over the coherent state projectors appears,
but it does \emph{not} generate an identity operator for all fermion
states, due to number conservation. Because the fermionic coherent
states are defined only for a subset of the Hilbert space, it is not
possible to obtain a continuous representation of unity.

Phase space representations for fermions \cite{Cahill_Galuber_fermions_1999}
can also be defined in terms of Grassmann coherent states \cite{Ohnuki_1978_CSFermiOp_PathInt,Cahill_Galuber_fermions_1999,Negele_Book_Qmanyparticlesystems,Blaizot_Book},
using anti-commuting Grassmann variables\cite{Berezin_book_SQMethod}.
As the Grassmann coherent states form a complete set, it is possible
in this case to obtain a resolution of unity. However, the use of
Grassmann variables means that the phase-space is defined over non-commuting
variables. These have an exponentially large matrix representation
in terms of standard complex variables. Consequently, they have no
efficient direct computational representation without requiring exponentially
large computational resources. In addition, while Grassmann coherent
states are mathematically well-defined, they do not correspond to
any physical state.

There is a close relationship between our work and the theory of random
matrices. The transformations we use correspond to the nonstandard
symmetry classes of random matrices. In this context, an important
consequence of our result is that it provides an exact solution to
a random matrix distribution of fermionic canonical ensembles. We
show that in the nonstandard symmetry classes, an even distribution
over matrix eigenvalues reduces a general canonical ensemble to the
identity operator. In other words, canonical ensembles in these symmetry
classes are nontrivial only if the random matrix distributions are
\emph{not} even functions of the eigenvalue. Nonstandard classes of
transformations can still have useful properties if the eigenvalue
distributions are not even functions - for example, if there is a
lower cutoff to the eigenvalue distribution. Similarly, our result
clearly does not exclude the use of ensembles that are not canonical.

This paper is organized as follows: In section \ref{sec:Gaussian-Operators}
we define the general Gaussian operators. In section \ref{sec:Symmetry-classes},
we define the symmetry classes that we use. In section \ref{sec:Resolution-of-unity}
we give the resolution of unity for the general Gaussian operators,
and in section \ref{sec:Identity-resolution-for-number-conserving}
we consider the case of the number-conserving Gaussian operators.
Section \ref{sec:Summary_Conclusion} gives a summary of our results,
conclusions and outlook. 

Additional results and integration identities are given in the Appendix.

\section{Fermionic gaussian operators\label{sec:Gaussian-Operators}}

We start by reviewing the known properties of fermionic Gaussian operators.
This will help to establish the notation, as well as to give results
that will be used in later parts of the paper. We consider a fermionic
system that can be decomposed into $M$ spatial or internal modes.
In some treatments of symmetry properties the space and spin indices
are treated separately, but this is not essential to our results.
We define $\hat{\bm{a}}$ as a vector of $M$ annihilation operators
and $\hat{\bm{a}}^{\dagger}$ as vector of $M$ creation operators.
As they are Fermi operators, $\hat{a}_{i}$ and $\hat{a}_{j}^{\dagger}$
obey the anticommutation relations:
\begin{equation}
\left\{ \hat{a}_{i},\hat{a}_{j}^{\dagger}\right\} =\delta_{ij},\qquad\left\{ \hat{a}_{i},\hat{a}_{j}\right\} =0.
\end{equation}
We define an extended vector of all $2M$ operators $\hat{\bm{\gamma}}=\left(\hat{\bm{a}},\hat{\bm{a}}^{\dagger}\right)$,
with an adjoint vector defined as $\hat{\bm{\gamma}}^{\dagger}=\left(\hat{\bm{a}}^{\dagger},\hat{\bm{a}}\right)=\left(\hat{a}_{1}^{\dagger},\ldots,\hat{a}_{M}^{\dagger},\hat{a}_{1},\ldots,\hat{a}_{M}\right)$.
In the remainder of this section, we will summarize properties of
the fermionic Gaussian operators to be used later.

\subsection{Group properties}

Following the original work of Balian and Brezin \cite{Balian_Brezin_Transformations},
we consider general linear transformations of fermions. These are
obtained from the fermionic Gaussian operator defined as an exponential
of a general quadratic form in Fermi operators:
\begin{equation}
\hat{G}_{B}\left(\mathbf{R}\right)=\exp\left[\frac{1}{2}\bm{\gamma}\mathbf{R}\bm{\gamma}\right],
\end{equation}
where $\mathbf{R}$ is a $2M\times2M$ antisymmetric complex matrix.
The group composition law for the general Gaussian operator is 
\begin{equation}
\hat{G}_{B}\left(\mathbf{R}\right)=\hat{G}_{B}\left(\mathbf{R}^{(1)}\right)\hat{G}_{B}\left(\mathbf{R}^{(2)}\right),\label{eq:Group_Comp_GGO}
\end{equation}
The value of $\mathbf{R}$ under this composition law is obtained
on defining:
\begin{equation}
\mathbf{H}=\bm{\sigma}\mathbf{R},
\end{equation}
where we have introduced a symmetric matrix: 
\begin{equation}
\bm{\sigma}=\left(\begin{array}{cc}
\mathbf{0}_{M} & \mathbf{I}_{M}\\
\mathbf{I}_{M} & \mathbf{0}_{M}
\end{array}\right).
\end{equation}
With this definition, the matrices follow a composition law for the
matrix parameters given by:
\begin{equation}
\exp\left(\mathbf{H}\right)=\exp\left(\mathbf{H}^{(1)}\right)\exp\left(\mathbf{H}^{(2)}\right).\label{eq:composition-law-general}
\end{equation}

This is clearly a Lie group. From the composition law, it has an inverse
and an identity, as well as being associative and differentiable.
It is equivalent to the $2M\times2M$ complex orthogonal Lie group,
which follows\cite{Balian_Brezin_Transformations} from the anti-symmetry
of $\mathbf{R}$.

\subsection{Hermitian sub-group}

The general Gaussian operator for fermions can also be usefully defined
in terms of the $\mathbf{H}$ matrix. Since this is the form most
commonly used in later work, we will use it here. With this definition,
the general un-normalized fermionic Gaussian operator is:
\begin{eqnarray}
\hat{G}\left(\mathbf{H}\right) & = & \exp\left[\frac{1}{2}\hat{\bm{\gamma}}^{\dagger}\mathbf{H}\hat{\bm{\gamma}}\right].\label{eq:GGO_ClassD}
\end{eqnarray}
 Here $\mathbf{H}$ is an $2M\times2M$ matrix, whose definition in
terms of an antisymmetric matrix $\mathbf{R}$ implies that it has
the decomposition
\begin{equation}
\mathbf{H}=\left(\begin{array}{cc}
\mathbf{h} & \bm{\Delta}\\
-\bm{\Delta^{*}} & -\mathbf{h}^{T}
\end{array}\right),\label{eq:H_Matrix}
\end{equation}
 where $\mathbf{h}$ and $\bm{\Delta}$ are $M\times M$ matrices.
Hence the quadratic term in the exponent can be rewritten as:
\begin{equation}
\hat{H}=\frac{1}{2}\hat{\bm{\gamma}}^{\dagger}\mathbf{H}\hat{\bm{\gamma}}=\frac{1}{2}\left(\hat{\bm{a}}^{\dagger}\mathbf{h}\hat{\bm{a}}-\hat{\bm{a}}\mathbf{h}^{T}\hat{\bm{a}}^{\dagger}+\hat{\bm{a}}^{\dagger}\Delta\hat{\bm{a}}^{\dagger}-\hat{\bm{a}}\Delta^{*}\hat{\bm{a}}\right).\label{eq:Hamiltonian-nonstandard}
\end{equation}

We note that, as well as being useful for linear transformations,
the same class of fermionic Gaussian operators can also be used to
expand a general fermionic density matrix in a positive phase-space
distribution\cite{Corney_PD_PRL2004_GQMC_ferm_bos,Corney_PD_PRB_2006_GPSR_fermions,Corney_PD_JPA_2006_GR_fermions}.
In general, this distribution includes both hermitian and non-hermitian
operators. 

In this form it is clear that if we wish to restrict the Gaussian
operators to be hermitian, we must require that $\mathbf{h}=\mathbf{h}^{\dagger}$,
and $\Delta=-\Delta^{T}$. This is the form we will use here to establish
the resolution of the fermionic identity operator. With this hermitian
restriction, the quadratic form has a clear physical identification.
It is simply the Bogoliubov-de Gennes Hamiltonian obtained from linearizing
the Hamiltonian for a superconductor. From this perspective, we see
that the Gaussian operator can have intrinsic coherence properties
that correspond to either a superconductor with $\Delta\neq0$, or
to a normal fluid with $\Delta=0$. Because it carries phase information,
the matrix $\Delta$ plays a similar role in fermion physics to the
bosonic coherent state amplitude, which appears in various approaches
to laser and superfluid theory.

\subsection{Number-conserving case\label{sub:Number-conserving-GO}}

Next we consider a subset of the general Gaussian operators with $\Delta=0$.
These are the \emph{number-conserving }fermionic Gaussian operators.
In this case, $\hat{\mathbf{H}}$ corresponds to the fermionic Hamiltonian
of a non-interacting Fermi gas in an arbitrary spin-dependent potential.
In general, these Gaussian operators form a subgroup of the complex
linear group, $GL(M,C)$. Balian and Brezin \cite{Balian_Brezin_Transformations}
found that these operators can be rewritten in terms of a single $M\times M$
matrix:
\begin{eqnarray}
\hat{G}_{N}\left(\mathbf{h}\right) & = & \exp\left[\hat{\bm{a}}^{\dagger}\mathbf{h}\hat{\bm{a}}-\frac{1}{2}Tr\left(\mathbf{h}\right)\right].
\end{eqnarray}

Using the matrix identity $\exp\left({\rm Tr}\left(\mathbf{h}\right)\right)=\det\left(\exp\left(\mathbf{h}\right)\right)$,
the number-conserving Gaussian operators can be written as:
\begin{equation}
\hat{G}_{N}\left(\mathbf{h}\right)=\frac{1}{\sqrt{\det\left(\exp\left(\mathbf{h}\right)\right)}}\exp\left[\hat{\bm{a}}^{\dagger}\mathbf{h}\hat{\bm{a}}\right].
\end{equation}
 The group composition law for these operators is given by:
\begin{equation}
\hat{G}_{N}\left(\mathbf{h}\right)=\hat{G}_{N}\left(\mathbf{h}^{(1)}\right)\hat{G}_{N}\left(\mathbf{h}^{(2)}\right),\label{eq:Group_Comp__ThermalGO}
\end{equation}
Here the composition law for $\mathbf{h}$ is obtained on matrix multiplication
of the variables $\mathbf{u}=e^{\mathbf{h}}$. The reason for this
is that multiplying the $\mathbf{u}^{(j)}$ matrices is equivalent
to multiplying the $\exp\left(\mathbf{H}^{(j)}\right)$ matrices defined
in Eq. (\ref{eq:composition-law-general}). Hence, we see that:
\begin{equation}
e^{\mathbf{h}}=e^{\mathbf{h}^{(1)}}e^{\mathbf{h}^{(2)}}.\label{eq:Composition_Law}
\end{equation}

We can equivalently express this Gaussian operator in terms of normally-ordered
parameters, using the following general mathematical identity for
an $M$-mode fermionic operator \cite{Fan_Operator_ordering_fermions}:
\begin{eqnarray}
\exp\left[\hat{\bm{a}}^{\dagger}\mathbf{h}\hat{\bm{a}}\right] & = & :\exp\left[\hat{\bm{a}}^{\dagger}\left[e^{\mathbf{h}}-\underline{\bm{\mathrm{I}}}\right]\hat{\bm{a}}\right]:\,.\label{eq:Mmode_ordering_fermions}
\end{eqnarray}

Therefore we obtain: 
\begin{eqnarray}
\hat{G}_{N}\left(\mathbf{h}\right) & = & e^{-\frac{1}{2}{\rm Tr}\left[\mathbf{h}\right]}:\exp\left[\hat{\bm{a}}^{\dagger}\left[e^{\mathbf{h}}-\underline{\bm{\mathrm{I}}}\right]\hat{\bm{a}}\right]:\,.\label{eq:un-norm_NumbCons_GO}
\end{eqnarray}

Hence, we notice that if we use normal ordering the simplest parameterization
is through the parameter $\mathbf{u}=e^{\mathbf{h}}$, which gives
a matrix-multiplication group composition law for these operators.
The quadratic term in the exponent can also be written in a Hamiltonian-like
form, in terms of a thermal or number-conserving Hamiltonian $\hat{\mathbf{H}}_{N}$:
\begin{equation}
\hat{H}_{N}=\hat{\bm{a}}^{\dagger}\mathbf{h}\hat{\bm{a}}-\frac{1}{2}{\rm Tr}\left(\mathbf{h}\right)=\frac{1}{2}\left(\hat{\bm{a}}^{\dagger}\mathbf{h}\hat{\bm{a}}-\hat{\bm{a}}\mathbf{h}^{T}\hat{\bm{a}}^{\dagger}\right)\,.
\end{equation}
We recognize that to obtain a hermitian Gaussian operator, it is necessary
to restrict $\mathbf{h}$ to the class of hermitian matrices.

\section{Symmetry classes and groups\label{sec:Symmetry-classes}}

The study of general symmetry groups for many-body systems originates
in the work of Wigner \cite{Wigner_1951,Wigner:1958} and Dyson \cite{Dyson:1962_JMP140,Dyson:1962_Threefold},
who studied the energy levels of complex many body systems such as
nuclei. In particular, Dyson \cite{Dyson:1962_JMP140,Dyson:1962_Threefold}
classified many-body systems depending on the symmetry properties
of the Hamiltonian. He classified Hamiltonians according to their
time-reversal and rotational invariance properties. 

In random matrix theory, this classification corresponds to three
random matrix models: the Gaussian unitary ensemble (GUE), the Gaussian
orthogonal ensemble (GOE), and the Gaussian symplectic ensemble (GSE)
\cite{Mehta_RM_book,Caselle_Magnea:2004_ReviewRMT_SymSpaces}. The
unitary ensemble applies to general systems without any invariance
under time reversal, the orthogonal ensemble is for systems with time-reversal
and rotational invariance, while the symplectic ensemble is for systems
with time-reversal invariance but no rotational invariance \cite{Mehta_RM_book,Caselle_Magnea:2004_ReviewRMT_SymSpaces}. 

Dyson's classification is based on number-conserving Hamiltonians.
If we consider the more general class of non-number-conserving Hamiltonians,
there are four additional symmetry classes that can be identified
depending on their time reversal symmetry and spin-rotation invariance
\cite{Altland_Zirnbauer:1997}. These four symmetry classes are Class
D, Class C, Class DIII and Class CI, which are related to Cartan's
classification of symmetric spaces. Class D corresponds to systems
with neither time-reversal symmetry nor spin-rotation symmetry. Class
DIII corresponds to systems with time-reversal symmetry but no spin-rotation
invariance. Class C is for systems with spin-rotation invariance and
no time-reversal invariance, while Class CI is for systems with spin-rotation
invariance and time-reversal invariance.

\subsection{Gaussian operators in nonstandard symmetry classes\label{sub:Gaussian_Operator_ClassD}}

The nonstandard symmetry classes are the general symmetry classes
of matrices that involve non-number-conserving Hamiltonians and particle-hole
symmetry. Here we focus mainly on the Class D symmetry defined by
Altland and Zirnbauer \cite{Altland_Zirnbauer:1997}, although our
results are applicable to any of the nonstandard symmetry classes.
Class D corresponds to the most general nonstandard symmetry class,
as it applies to cases which do not have time-reversal or spin-rotation
symmetry. In this case the only required properties of the matrices
$\mathbf{h}$ and $\bm{\Delta}$, defined in Eq. (\ref{eq:H_Matrix}),
are the hermiticity of $\mathbf{h}$ and skew-symmetry of $\bm{\Delta}$.
The $2M\times2M$ $\mathbf{H}$ matrix is therefore hermitian, as
discussed above. We start by summarizing group theoretic results of
Altland and Zirnbauer, then extend these to obtain new results for
distributions over canonical ensembles.

Hermitian matrices do not themselves form a group because they do
not close under commutation. However, the anti-hermitian matrices
form a Lie group, so it is useful to define an anti-hermitian matrix
$\mathbf{X}$, as $\mathbf{X}=i\mathbf{H}$. Hence the conditions
on $\mathbf{h}$ and $\bm{\Delta}$ can be expressed in terms of the
anti-hermitian matrix $\mathbf{X}$ as:
\begin{equation}
-\mathbf{X}^{\dagger}=\mathbf{X}=-\bm{\Sigma}_{X}\mathbf{X}^{T}\bm{\Sigma_{x},}\label{eq:X_matrix}
\end{equation}
where the matrix $\bm{\Sigma}_{X}$ is defined as:
\[
\bm{\Sigma}_{x}=\left(\begin{array}{cc}
\mathbf{0}_{M} & \mathbf{I}_{M}\\
\mathbf{I}_{M} & \mathbf{0}_{M}
\end{array}\right).
\]
 The Lie algebra of the matrices defined in Eq. (\ref{eq:X_matrix})
is isomorphic to the $so(2M)$ algebra\cite{Altland_Zirnbauer:1997}.
Since $\mathbf{X}$ belongs to a Lie algebra, it can be diagonalized:
\begin{equation}
\mathbf{X}=\mathbf{U}^{-1}\widetilde{\bm{\lambda}}\mathbf{U}.\label{eq:Diagonalization_X}
\end{equation}

Here $\mathbf{U}$ is an element of a Lie group which is isomorphic
to $SO(2M)$. The diagonalization of Eq. (\ref{eq:Diagonalization_X})
is equivalent to decomposing the $\mathbf{X}$ matrix in matrix polar
coordinates. The radial part corresponds to the $\widetilde{\bm{\lambda}}$
matrix, $\widetilde{\bm{\lambda}}={\rm diag}\left(i\bm{\lambda},-i\bm{\lambda}\right)$,
where $\bm{\lambda}$ is an $M\times M$ diagonal matrix of eigenvalues
$\bm{\lambda}=\left\{ \lambda_{1},\ldots,\lambda_{M}\right\} $, and
the angular part corresponds to the $\mathbf{U}$ matrix.

\subsection{Diagonal form}

Let us now consider the exponent of the Gaussian operator defined
in Eq. (\ref{eq:GGO_ClassD}). After diagonalization, we can write
this operator expression in terms of the matrix $\mathbf{X}$ as:
\begin{eqnarray*}
\hat{H}=\frac{1}{2}\hat{\bm{\gamma}}^{\dagger}\mathbf{H}\hat{\bm{\gamma}} & = & \frac{1}{2}\hat{\bm{\gamma}}^{\dagger}\left(-i\mathbf{X}\right)\hat{\bm{\gamma}}\\
 & = & \frac{-i}{2}\left(\hat{\bm{a}}^{\dagger},\hat{\bm{a}}\right)\left(\mathbf{U}^{-1}\widetilde{\bm{\lambda}}\mathbf{U}\right)\left(\begin{array}{c}
\hat{\bm{a}}\\
\hat{\bm{a}}^{\dagger}
\end{array}\right).
\end{eqnarray*}

Since the canonical anticommutation relations are invariant under
this transformation, we can define new fermionic operators 
\begin{equation}
\left(\begin{array}{c}
\hat{\bm{b}}\\
\hat{\bm{b}}^{\dagger}
\end{array}\right)=\mathbf{U}\left(\begin{array}{c}
\hat{\bm{a}}\\
\hat{\bm{a}}^{\dagger}
\end{array}\right),\label{eq:Unitary_transf_operator}
\end{equation}
 therefore, we can show that:
\[
\hat{G}\left(\mathbf{H}\right)=\exp\left[\frac{-i}{2}\left(\hat{\bm{b}}^{\dagger},\hat{\bm{b}}\right)\left(\begin{array}{cc}
i\bm{\lambda} & \mathbf{0}\\
\mathbf{0} & -i\bm{\lambda}
\end{array}\right)\left(\begin{array}{c}
\hat{\bm{b}}\\
\hat{\bm{b}}^{\dagger}
\end{array}\right)\right]\,,
\]
and hence using anti-commutation relations it follows that: 
\begin{equation}
\hat{G}\left(\bm{\lambda}\right)=\exp\left[\bm{\lambda}\hat{\mathbf{b}}^{\dagger}\hat{\mathbf{b}}-\frac{1}{2}{\rm Tr}\left[\bm{\lambda}\right]\right].\label{eq:GO_eigenv_noOrder}
\end{equation}
We can write Eq. (\ref{eq:GO_eigenv_noOrder}) in normally ordered
form using Eq. (\ref{eq:Mmode_ordering_fermions}), hence:
\begin{equation}
\hat{G}\left(\bm{\lambda}\right)=e^{-\frac{1}{2}{\rm Tr\left[\bm{\lambda}\right]}}:\exp\left[\hat{\mathbf{b}}_{i}^{\dagger}\left(e^{\bm{\lambda}}-1\right)_{i}\hat{\mathbf{b}}_{i}\right]:\,.\label{eq:GO_eigenv_NormOrd}
\end{equation}

\subsection{Elementary properties }

We now wish to use this diagonal form to prove some elementary properties
of these Gaussian operators.

\subsubsection{Positivity}

The Gaussian operator defined in Eq. (\ref{eq:GO_eigenv_NormOrd})
is defined in terms of real eigenvalues, $\lambda_{j}\in\left(-\infty,\infty\right),$
so that $e^{\lambda_{j}}\in\left(0,\infty\right)$. Hence, if we consider
the normally ordered form we notice that 
\begin{eqnarray}
\hat{G}\left(\bm{\lambda}\right) & = & e^{-\frac{1}{2}{\rm Tr\left[\bm{\lambda}\right]}}:\exp\left[\hat{\mathbf{b}}_{i}^{\dagger}\left(e^{\bm{\lambda}}-1\right)_{i}\hat{\mathbf{b}}_{i}\right]:\nonumber \\
 & = & \prod_{i}e^{-\lambda_{i}/2}\left(1+\hat{\mathbf{b}}_{i}^{\dagger}\left(e^{\bm{\lambda}}-1\right)_{i}\hat{\mathbf{b}}_{i}\right)\ge0.
\end{eqnarray}

Therefore the Gaussian operators with hermitian \textbf{H} matrices
are themselves hermitian, positive definite operators in the fermionic
Hilbert space.

\subsubsection{Normalization}

The Gaussian operators defined in Eq. (\ref{eq:GO_eigenv_NormOrd})
can be normalized to obtain operators $\hat{\Lambda}(\mathbf{H})$,
such that ${\rm Tr}\left[\hat{\Lambda}(\mathbf{H})\right]=1$. We
first consider the trace of a single-mode case, which is given by:
\begin{equation}
{\rm Tr}\left[:\exp\left[\hat{b}^{\dagger}\left(e^{\lambda}-1\right)\hat{b}\right]:\right]={\rm Tr}\left[1+\hat{b}^{\dagger}\hat{b}\left(e^{\lambda}-1\right)\right]=1+e^{\lambda}.
\end{equation}
On the other hand, including the exponential factor obtained during
normal-ordering, 
\begin{eqnarray}
{\rm Tr}\left[e^{\hat{b}^{\dagger}\lambda\hat{b}-\frac{\lambda}{2}}\right] & = & \left(1+e^{\lambda}\right)e^{-\frac{\lambda}{2}}=2\cosh\left(\frac{\lambda}{2}\right).
\end{eqnarray}
Then for the single mode case we can introduce a normalized Gaussian
operator $\hat{\Lambda}(\lambda)$ such that $Tr\left[\hat{\Lambda}(\lambda)\right]=1$,
in the form $\hat{\Lambda}(\lambda)=\exp\left[\hat{b}^{\dagger}\lambda\hat{b}-\frac{\lambda}{2}\right]/\left[2\cosh\left(\lambda/2\right)\right].$

For the general $M$-mode case we can transform back to the original
fermionic operators, using invariance of the determinant under unitary
transformations to obtain:
\begin{equation}
\hat{\Lambda}(\mathbf{H})=\frac{e^{\frac{1}{2}\hat{\gamma}^{\dagger}\mathbf{H}\hat{\gamma}}}{\det\left[2\cosh\left(\mathbf{H}/2\right)\right]}.\label{eq:Norm_GO}
\end{equation}
We note that this normalized form is identical to that used in Gaussian
phase-space representations \cite{Corney_PD_PRL2004_GQMC_ferm_bos,Corney_PD_JPA_2006_GR_fermions,Corney_PD_PRB_2006_GPSR_fermions}.

\section{General resolution of unity\label{sec:Resolution-of-unity}}

\subsection{General Gaussian Operators }

We wish to prove that the resolution of unity for the normalized hermitian
Gaussian operator, $\hat{\Lambda}\left(\mathbf{H}\right)$, defined
in Eq. (\ref{eq:Norm_GO}) is given by:
\begin{eqnarray}
\int d\mathbf{H}\hat{\Lambda}\left(\mathbf{H}\right)P\left(\mathbf{H}^{2}\right) & = & \hat{I}.\label{eq:Resolution_unity_GO}
\end{eqnarray}

Here $\hat{I}$ is the fermionic identity operator, $d\mathbf{H}$
is the measure over hermitian matrices which conserves the nonstandard
group symmetries of any of the four different symmetry classes defined
by Altland and Zirnbauer \cite{Altland_Zirnbauer:1997}. Here, $P\left(\mathbf{H}^{2}\right)$
is a normalizable positive function of $\mathbf{H}^{2}$ that is invariant
under the transformation $\mathbf{U}$, where $\mathbf{U}$ depends
on the symmetry class. On taking a trace of the equation, we note
from Eq (\ref{eq:Resolution_unity_GO}), that the normalization of
this function must satisfy: $2^{-M}\int d\mathbf{H}P\left(\mathbf{H}^{2}\right)=1.$

Although our result is general, for definiteness we consider two options
for the normalization function:
\begin{eqnarray}
P^{(1)} & = & C^{(1)}\det\left[1+\mathbf{H}^{2}\right]^{-p},\label{eq:Norm1_det}\\
P^{(2)} & = & C^{(2)}\exp\left[-p{\rm Tr}\left[\mathbf{H}^{2}\right]\right].\label{eq:Norm2_Exp}
\end{eqnarray}

\subsection{Matrix polar coordinates}

We now wish to use a unitary transformation to reduce the integral
over matrix elements of $\mathbf{H}$ to matrix polar coordinates.
The angular variables correspond to unitary transformations, while
radial variables correspond to eigenvalues. The Jacobian for the transformation
from the Cartesian coordinates $\mathbf{H}=-i\mathbf{X}$ to polar
coordinates $\left(\bm{\lambda},\mathbf{U}\right)$, is given by diagonalizing
the matrix $\mathbf{H}$ defined in Eq. (\ref{eq:H_Matrix}) , so
that\cite{Altland_Zirnbauer:1997}
\begin{eqnarray}
d\mathbf{H} & = & \left(\mathbf{U}^{\dagger}d\mathbf{U}\right)\Delta^{\beta}\left(\bm{\lambda}^{2}\right)d\mathbf{\bm{\lambda}}\prod_{j}\left|\lambda_{j}\right|^{\alpha}\,.\label{eq:Jacobian}
\end{eqnarray}
Here $d\mathbf{\bm{\lambda}}=\prod_{j=1}^{M}d\lambda_{j}$, $\mathbf{U}$
is the transformation that diagonalizes the matrix $\mathbf{H}$ for
the symmetry class under consideration, and $\Delta\left(\bm{\lambda}\right)$
is the Vandermonde determinant defined as:
\begin{equation}
\Delta(\mathbf{\mathbf{\bm{\lambda}}})=\Delta(\lambda_{1},\ldots,\lambda_{M})=\prod_{1\le i<j\leq M}\left(\lambda_{i}-\lambda_{j}\right)=\det\left[\lambda_{i}^{j-1}\right].\label{eq:Vandermonde_determinant}
\end{equation}
The indices $\alpha$ and $\beta$ of Eq. (\ref{eq:Jacobian}), depend
on the underlying nonstandard symmetry class. Their values are given
in Table \ref{tab:Indices_Symmetry_clases_AZ}.

\begin{table}
\begin{centering}
\caption{Indices of the different symmetry classes.\label{tab:Indices_Symmetry_clases_AZ}}

\par\end{centering}

\centering{}%
\begin{tabular}{|c|c|c|}
\hline 
Class & $\beta$ & $\alpha$\tabularnewline
\hline 
\hline 
D & $2$ & $0$\tabularnewline
\hline 
C & $2$ & $2$\tabularnewline
\hline 
DIII & $4$ & $1$\tabularnewline
\hline 
C1 & $1$ & $1$\tabularnewline
\hline 
\end{tabular}
\end{table}

Eq. (\ref{eq:Resolution_unity_GO}) in polar coordinates can now be
written, after unitary transformation to diagonal operator form, as:
\begin{eqnarray}
\int d\mathbf{H}\hat{\Lambda}\left(\mathbf{H}\right)P\left(\mathbf{H}^{2}\right) & = & \int\left(\mathbf{U}^{\dagger}d\mathbf{U}\right)\int P\left(\bm{\lambda}^{2}\right)\Delta^{\beta}\left(\bm{\lambda}^{2}\right)d\mathbf{\bm{\lambda}}\times\nonumber \\
 &  & \times\prod_{j=1}^{M}\left[\frac{\left|\lambda_{j}\right|^{\alpha}e^{-\frac{1}{2}{\rm \lambda_{j}}}:\exp\left[\hat{b}_{j}^{\dagger}\left(e^{\bm{\lambda}}-1\right)_{j}\hat{b}_{j}\right]:}{\left[2\cosh\left(\lambda_{j}/2\right)\right]}\right].\label{eq:Int1}
\end{eqnarray}
Just as in the previous section, we have defined $\hat{b}_{j}$ as
a function of the unitary transformation $\mathbf{U}$, according
to Eq (\ref{eq:Unitary_transf_operator}).

\subsection{Evaluation of integrals}

We next wish to show that the result of the integral over radial variables
$\bm{\lambda}$ is a constant, independent of the transformation $\mathbf{U}$.
In order to evaluate the integral over the radial variables $\bm{\lambda}$,
we notice that the expansion of the operator term can be expressed
as:
\begin{eqnarray}
\prod_{j=1}^{M}e^{-\frac{1}{2}{\rm \lambda_{j}}}:\exp\left[\hat{\mathbf{b}}_{j}^{\dagger}\left(e^{\bm{\lambda}}-1\right)_{j}\hat{\mathbf{b}}_{j}\right]: & = & \prod_{j=1}^{M}e^{-\frac{1}{2}{\rm \lambda_{j}}}\left(1+\hat{b}_{j}^{\dagger}\left(e^{\bm{\lambda}}-1\right)_{j}\hat{b}_{j}\right)\nonumber \\
 & = & \prod_{j=1}^{M}\left(e^{-\frac{1}{2}{\rm \lambda_{j}}}+2\sinh\left(\frac{\lambda_{j}}{2}\right)\hat{b}_{j}^{\dagger}\hat{b}_{j}\right).\label{eq:Operator_term}
\end{eqnarray}
 Using this result to expand Eq. (\ref{eq:Int1}) we obtain:
\begin{eqnarray}
\int d\mathbf{\bm{\lambda}}\Delta^{\beta}\left(\bm{\lambda}^{2}\right)\hat{\Lambda}\left(\bm{\lambda}\right)P\left(\bm{\lambda}^{2}\right)\prod_{j=1}^{M}\left|\lambda_{j}\right|^{\alpha} & = & \int d\mathbf{\bm{\lambda}}P\left(\bm{\lambda}^{2}\right)\Delta^{\beta}\left(\bm{\lambda}^{2}\right)\prod_{j=1}^{M}\left[\frac{\left|\lambda_{j}\right|^{\alpha}e^{-\frac{1}{2}{\rm \lambda_{j}}}}{\left[2\cosh\left(\lambda_{j}/2\right)\right]}\right]\nonumber \\
 & + & \int d\mathbf{\bm{\lambda}}P\left(\bm{\lambda}^{2}\right)\Delta^{\beta}\left(\bm{\lambda}^{2}\right)\prod_{j=1}^{M}\left[\left|\lambda_{j}\right|^{\alpha}\tanh\left(\frac{\lambda_{j}}{2}\right)\hat{b}_{j}^{\dagger}\hat{b}_{j}\right].\qquad\label{eq:Int1_ResUnity}
\end{eqnarray}
 Here we see that the integral over the operator terms $\hat{b}_{i}^{\dagger}\hat{b}_{i}$
has $\tanh$ terms which are all odd in the $\lambda_{j}$ variable,
while every other term is an even function of $\lambda_{j}$. Therefore,
from the parity of these functions, all the terms of the second integral
in Eq. (\ref{eq:Int1_ResUnity}) vanish on integration over $\lambda_{j}$. 

Since the operator terms are the only terms that depend on the unitary
transformation $\mathbf{U}$, it follows that the integral over the
angular part is a constant, given by the relevant angular volume,
which we will denote by $C^{\mathbf{U}}$. The value of $C^{\mathbf{U}}$
depends on the corresponding non-standard symmetry class under consideration,
which defines the transformation $\mathbf{U}$. The result of the
integration over the eigenvalues then determines the normalization
factor.

Using the standard identity, $\exp\left(\lambda_{j}/2\right)=\cosh\left(\lambda_{j}/2\right)-\sinh\left(\lambda_{j}/2\right)$,
Eq. (\ref{eq:Int1_ResUnity}) can be written as:
\begin{eqnarray}
\int d\mathbf{H}\hat{\Lambda}\left(\mathbf{H}\right)P\left(\mathbf{H}^{2}\right) & = & C^{\mathbf{U}}2^{-M}\int d\mathbf{\bm{\lambda}}P\left(\bm{\lambda}^{2}\right)\Delta^{\beta}\left(\bm{\lambda}^{2}\right)\prod_{j=1}^{M}\left|\lambda_{j}\right|^{\alpha}\left[1-\tanh\left(\frac{\lambda_{j}}{2}\right)\right]\nonumber \\
 & = & 2^{-M}C^{\mathbf{U}}\int_{-\infty}^{\infty}d\mathbf{\bm{\lambda}}P\left(\bm{\lambda}^{2}\right)\Delta^{\beta}\left(\bm{\lambda}^{2}\right)\prod_{j=1}^{M}\left|\lambda_{j}\right|^{\alpha}.\label{eq:Int3_ResUnity}
\end{eqnarray}
In order to obtain Eq. (\ref{eq:Int3_ResUnity}), we have used the
result that the integral over the $\tanh\left(\lambda_{j}/2\right)$
terms vanishes, because clearly $\tanh\left(\lambda_{j}/2\right)$
is an odd function of $\lambda_{j}$ while the other terms are even.
Next, we recall that the original definition of $P$ was such that
it was normalized. Therefore, using matrix polar coordinates to evaluate
Eq (\ref{eq:Resolution_unity_GO}), we find that:
\begin{eqnarray}
1 & = & 2^{-M}\int d\mathbf{H}P\left(\mathbf{H}^{2}\right)\nonumber \\
 & = & 2^{-M}C^{\mathbf{U}}\int_{-\infty}^{\infty}d\mathbf{\bm{\lambda}}P\left(\bm{\lambda}^{2}\right)\Delta^{\beta}\left(\bm{\lambda}^{2}\right)\prod_{j=1}^{M}\left|\lambda_{j}\right|^{\alpha}.
\end{eqnarray}

For any non-standard symmetry class it is therefore possible to express
the resolution of unity as:
\begin{eqnarray}
\hat{I} & = & \int d\mathbf{H}\hat{\Lambda}\left(\mathbf{H}\right)P\left(\mathbf{H}^{2}\right)\nonumber \\
 & = & 2^{-M}C^{\mathbf{U}}\int_{-\infty}^{\infty}d\mathbf{\bm{\lambda}}P\left(\bm{\lambda}^{2}\right)\Delta^{\beta}\left(\bm{\lambda}^{2}\right)\prod_{j=1}^{M}\left|\lambda_{j}\right|^{\alpha}.\label{eq:ResUnity_NonstandardSym}
\end{eqnarray}
 The value of both the angular volume $C^{\mathbf{U}}$ and the radial
integral will depend on the corresponding non-standard symmetry class
under consideration. For the radial part, the integral will also depend
on the different choices of the normalization function $P(\bm{\lambda}^{2})$.
We notice that as long as we can perform the integration over the
angular and radial part, then it is possible to obtain a resolution
of unity for any of the non-standard symmetry classes. An explicit
result for these integrals in the case of class D symmetry is given
next.

We note there is an important consequence of this result in random
matrix theory. Suppose we consider a statistical random matrix mixture
of finite temperature canonical ensembles given by:
\begin{equation}
\rho=\int d\mathbf{H}\, P\left(\mathbf{H}^{2}\right)\exp\left[-\beta\hat{H}\right]\propto\hat{I}\,,
\end{equation}
where $\hat{H}$ is the linearized Bogoliubov-de Genne Hamiltonian
given by Eq (\ref{eq:Hamiltonian-nonstandard}), and $d\mathbf{H}$
is a measure over one of the four nonstandard symmetry classes. It
follows that any correlation function or moment evaluated in this
ensemble is simply an average over the identity operator, independent
of temperature, symmetry class or the details of the distribution
$P\left(\mathbf{H}^{2}\right)$.

\subsection{Symmetry Class D}

In order to give the values of the angular volume and radial integrals
 defined in Eq. (\ref{eq:ResUnity_NonstandardSym}), we will consider
the most general symmetry class D. That is, we are going to integrate
over the transformations described in Sec. \ref{sub:Gaussian_Operator_ClassD}.
We note, however, that similar results in any of the non-standard
symmetry classes can be found. The main differences are in the values
of the integration volumes, and in convergence properties which depend
on the details of the Jacobian and matrix polar coordinates. 

For symmetry class D, which is the largest symmetry class, the values
of the indices $\alpha$ and $\beta$, given in Table \ref{tab:Indices_Symmetry_clases_AZ},
are $0$ and $2$ respectively\cite{Altland_Zirnbauer:1997}. In this
case the value of the integral over the angular part, $C^{\mathbf{U}}=\mathbf{U}^{\dagger}d\mathbf{U}$
is given below. Details are given in the Appendix:
\begin{eqnarray}
C^{\mathbf{U}}=\int\left(\mathbf{U}^{\dagger}d\mathbf{U}\right) & =\frac{\pi^{M\left(M-\frac{1}{2}\right)}}{2^{M\left(M-1\right)}} & \prod_{j=0}^{M-1}\frac{1}{\Gamma\left(2+j\right)\Gamma\left(j+\frac{1}{2}\right)}.\label{eq:Angular_Integral}
\end{eqnarray}

Hence, for the non-standard symmetry class D, the integral  defined
in Eq. (\ref{eq:ResUnity_NonstandardSym}) is:
\begin{eqnarray}
\hat{I} & = & \int d\mathbf{H}\hat{\Lambda}\left(\mathbf{H}\right)P\left(\mathbf{H}^{2}\right)\nonumber \\
 & = & 2^{-M}C^{\mathbf{U}}\int_{-\infty}^{\infty}P\left(\bm{\lambda}^{2}\right)\Delta^{2}\left(\bm{\lambda}^{2}\right)d\mathbf{\bm{\lambda}}.
\end{eqnarray}

\subsubsection{Determinant normalization}

In order to evaluate the integral over the radial variable $\bm{\lambda}$,
we have to consider one of the two options for the normalization $P\left(\bm{\lambda}^{2}\right)$.
We first consider the normalization by the determinant given in Eq.
(\ref{eq:Norm1_det}). Hence the integral over the variables $\bm{\lambda}$
is:
\begin{eqnarray}
\int_{-\infty}^{\infty}d\mathbf{\bm{\lambda}}\Delta^{2}\left(\bm{\lambda}^{2}\right)P^{(1)}\left(\bm{\lambda}^{2}\right) & = & C^{(1)}\int_{-\infty}^{\infty}d\mathbf{\bm{\lambda}}\Delta^{2}\left(\bm{\lambda}^{2}\right)\prod_{j=1}^{M}\left(1+\lambda_{j}^{2}\right)^{-2p}.\label{eq:Selberg_Int}
\end{eqnarray}

The integral of the right hand side of Eq. (\ref{eq:Selberg_Int})
is known as Selberg's Integral \cite{Mehta_RM_book}. Selberg's formula
is given below. It is valid for integer $n$ and complex $\alpha,$
$\beta$, $\gamma$ with ${\rm Re}\alpha>0$, ${\rm Re}\beta>0$ and
${\rm Re}\gamma>-{\rm min}\left(\frac{1}{n},\frac{{\rm Re}\alpha}{n-1},\frac{{\rm Re}\beta}{n-1}\right)$:
\begin{eqnarray}
I\left(\alpha,\beta,\gamma,n\right) & = & \int_{0}^{\infty}\ldots\int_{0}^{\infty}\left|\Delta\left(\mathbf{x}\right)\right|^{2\gamma}\prod_{j=1}^{n}x_{j}^{\alpha-1}\left(1+x_{j}\right)^{-\alpha-\beta-2\gamma\left(n-1\right)}dx_{j}\nonumber \\
 & = & \prod_{j=0}^{n-1}\frac{\Gamma\left(1+\gamma+j\gamma\right)\Gamma\left(\alpha+j\gamma\right)\Gamma\left(\beta+j\gamma\right)}{\Gamma\left(1+\gamma\right)\Gamma\left(\alpha+\beta+\left(n+j-1\right)\gamma\right)}.\label{eq:Selberg_Integral}
\end{eqnarray}
Defining $\upsilon_{j}=\lambda_{j}^{2},$ $d\upsilon=2\lambda d\lambda_{j},$
$d\lambda_{j}=\frac{1}{2}\upsilon^{-\frac{1}{2}}d\upsilon,$ and setting
$n=M,$ $\alpha=1/2$, $\gamma=1$ and $-\alpha-\beta-2\gamma\left(n-1\right)=-2p,$
so that $\beta=-2M+2p+3/2,$ we require $p>M-\frac{3}{4}$, which
gives the result:
\begin{eqnarray}
\int_{-\infty}^{\infty}d\mathbf{\bm{\lambda}}\Delta^{2}\left(\bm{\lambda}^{2}\right)\prod_{j=1}^{M}\left(1+\lambda_{j}^{2}\right)^{-2p} & = & \prod_{j=0}^{M-1}\frac{\Gamma\left(2+j\right)\Gamma\left(\frac{1}{2}+j\right)\Gamma\left(-2M+2p+3/2+j\right)}{\Gamma\left(-M+2p+j+1\right)}.
\end{eqnarray}

Therefore, one fermionic resolution of unity in the most general symmetry
class D is given by:
\begin{equation}
\hat{I}=\int d\mathbf{H}\hat{\Lambda}\left(\mathbf{H}\right)P^{(1)}\left(\mathbf{H}^{2}\right)\,,
\end{equation}
where the normalization constant $C^{(1)}$ is given by:

\begin{equation}
C^{(1)}=\frac{2^{M^{2}}}{\pi^{M\left(M-\frac{1}{2}\right)}}\prod_{j=0}^{M-1}\frac{\Gamma\left(-M+2p+j+1\right)}{\Gamma\left(-2M+2p+j+\frac{3}{2}\right)}.
\end{equation}

\subsubsection{Gaussian normalization}

Now we consider the second option for the normalization function,
given in Eq. (\ref{eq:Norm2_Exp}), where we normalize the distribution
by a Gaussian c-number function of the eigenvalues, as often used
in random matrix theory. Since this normalization is another even
function, the integral over Gaussian operators reduces to a term proportional
to the identity operator, as before. We now wish to evaluate the normalization
constant, to obtain a resolution of unity. In this case the integral
over the radial part is:
\begin{eqnarray}
\int_{-\infty}^{\infty}d\mathbf{\bm{\lambda}}\Delta^{2}\left(\bm{\lambda}^{2}\right)P^{(2)}\left(\bm{\lambda}^{2}\right) & = & C^{(2)}\int_{-\infty}^{\infty}d\mathbf{\bm{\lambda}}\Delta^{2}\left(\bm{\lambda}^{2}\right)e^{-2p\sum_{j}\lambda_{j}^{2}}.\label{eq:Selberg_Int_Laguerre}
\end{eqnarray}

The integral of Eq. (\ref{eq:Selberg_Int_Laguerre}) is a Selberg
type integral related to the Laguerre polynomials \cite{Mehta_RM_book}:
\[
\int_{-\infty}^{\infty}\ldots\int_{-\infty}^{\infty}\left|\Delta\left(x^{2}\right)\right|^{2\gamma}\prod_{j=1}^{n}\left|x_{j}\right|^{2\tilde{\alpha}-1}\exp\left(-\frac{x_{j}^{2}}{2}\right)d\mathbf{x}=2^{\tilde{\alpha}n+\gamma n\left(n-1\right)}\prod_{j=1}^{n}\frac{\Gamma\left(1+j\gamma\right)\Gamma\left(\tilde{\alpha}+\gamma\left(j-1\right)\right)}{\Gamma\left(1+\gamma\right)}.
\]

We set $\tilde{\alpha}=1/2$, $\gamma=1$, $n=M$ and $x_{j}^{2}=4p\lambda_{j}$.
Hence we obtain:
\begin{eqnarray*}
\int_{-\infty}^{\infty}P^{(2)}\left(\bm{\lambda}^{2}\right)\Delta^{2}\left(\bm{\lambda}^{2}\right)d\mathbf{\bm{\lambda}} & = & \left(2p\right)^{-M\left(M-\frac{1}{2}\right)}\prod_{j=1}^{M}\Gamma\left(1+j\right)\Gamma\left(j-\frac{1}{2}\right).
\end{eqnarray*}
 In this case the resolution of unity is
\begin{equation}
\hat{I}=\int d\mathbf{H}\hat{\Lambda}\left(\mathbf{H}\right)P^{(2)}\left(\mathbf{H}^{2}\right)\,,
\end{equation}
where the normalizing constant is given by

\begin{equation}
C^{(2)}=2^{M^{2}}\left(\frac{2p}{\pi}\right)^{M(M-1/2)}.
\end{equation}

\section{number-conserving Gaussian operators\label{sec:Identity-resolution-for-number-conserving}}

By analogy to the general Gaussian operators, we wish to investigate
if there is a similar expression for the fermionic identity operator
in terms of the number-conserving Gaussian operators. These operators,
$\hat{G}_{N}\left(\mathbf{h}\right),$ defined in Eq. (\ref{eq:un-norm_NumbCons_GO})
can be normalized so that ${\rm Tr}\left[\hat{G}_{N}\left(\mathbf{h}\right)\right]=1$.
The normalized number-conserving Gaussian operator $\hat{\Lambda}_{N}\left(\mathbf{h}\right)$
is:
\begin{equation}
\hat{\Lambda}_{N}(\mathbf{h})=\frac{e^{-\frac{1}{2}{\rm Tr}\left(\mathbf{h}\right)}}{\det\left(2\cosh\left(\mathbf{h}/2\right)\right)}:\exp\left[\hat{\bm{a}}^{\dagger}\left[e^{\mathbf{h}}-\underline{\bm{\mathrm{I}}}\right]\hat{\bm{a}}\right]:\,.\label{eq:Norm_number-cons_GO}
\end{equation}
 In section \ref{sub:Number-conserving-GO}, we introduced the variable
$\mathbf{u}=e^{\mathbf{h}}$. We notice that if we define the variable
$\mathbf{u}$ as $\mathbf{u}=\tilde{\mathbf{n}}^{-T}-\underline{\bm{\mathrm{I}}},$
we obtain the expression for the normalized number-conserving Gaussian
operator defined in terms of the stochastic Green's functions\cite{Corney_PD_JPA_2006_GR_fermions,Corney_PD_PRB_2006_GPSR_fermions}
for particles, $\mathbf{n}$ and holes $\tilde{\mathbf{n}}=\underline{\bm{\mathrm{I}}}-\mathbf{n}$:
\begin{equation}
\hat{\Lambda}_{N}(\mathbf{h})=\det\left[\tilde{\mathbf{n}}\right]:\exp\left[\hat{\bm{a}}^{\dagger}\left[\tilde{\mathbf{n}}^{-1}-2\bm{\mathrm{I}}\right]^{T}\hat{\bm{a}}\right]:\,.\label{eq:un-norm_GO_n}
\end{equation}

\subsection{Matrix polar coordinates}

We wish to investigate if there is a normalization factor ${\cal N}(\mathbf{h})$
that generates a resolution of unity for the normalized number-conserving
Gaussian operators, $\hat{\Lambda}_{N}\left(\mathbf{h}\right)$ defined
in Eq. (\ref{eq:Norm_number-cons_GO}), of the form:
\begin{equation}
\int\hat{\Lambda}_{N}\left(\mathbf{h}\right){\cal N}(\mathbf{h})d\mathbf{h}=\hat{I}.\label{eq:ResUnity_NCGO}
\end{equation}
Here $d\mathbf{h}$ is an integration measure over the hermitian matrices
$\mathbf{h}$ and ${\cal N}(\mathbf{h})$ is a normalization function.
This normalization function is defined in order to ensure the convergence
of the integral. We can consider, for example, the following option
for the normalization, in analogy with the previous case:
\begin{eqnarray}
{\cal N}\left(\mathbf{h}\right) & = & \frac{C}{\det\left[\cosh\left(\mathbf{h}/2\right)\right]^{p}}\,,\label{eq:Normalization_GO}
\end{eqnarray}
where $C$ is a normalization constant and we require $p>1$. 

Hermitian matrices can be decomposed in polar coordinates as\cite{Hua_Book_harmonic_analysis}
$\mathbf{h}=\mathbf{U}\bm{\lambda}\mathbf{U}^{\dagger},$ where $\mathbf{U}$
is an unitary matrix and corresponds to the angular coordinates, while
$\bm{\lambda}$ corresponds to the radial coordinates and is a diagonal
matrix, $\bm{\lambda}=\left\{ \lambda_{1},\ldots,\lambda_{M}\right\} $.
The Jacobian of the transformation from the cartesian coordinates
$\mathbf{h}$ to polar coordinates $\left(\lambda,\mathbf{U}\right)$
is given by \cite{Itzykson:1980}:
\begin{equation}
d\mathbf{h}=d\mathbf{U}\Delta^{2}(\mathbf{\mathbf{\bm{\lambda}}})\prod_{i}d\lambda_{i}.\label{eq:Hermitian_measure_IZ}
\end{equation}
Here $d\mathbf{U}$ is the normalized Haar measure over unitary matrices
and $\Delta(\mathbf{\mathbf{\bm{\lambda}}})$ is the Vandermonde determinant
defined in Eq. (\ref{eq:Vandermonde_determinant}).

\subsection{Evaluation of integrals}

Hence, on diagonalizing the Gaussian operator:
\begin{eqnarray}
\hat{\Lambda}_{N}\left(\mathbf{h}\right) & = & :\exp\left[\hat{\bm{b}}^{\dagger}\left(e^{\mathbf{\bm{\lambda}}}-\mathbf{I}\right)\hat{\bm{b}}\right]:\prod_{j=1}^{M}\frac{e^{-\frac{1}{2}{\rm \lambda_{j}}}}{2\cosh\left(\lambda_{j}/2\right)}\,,
\end{eqnarray}
 where we have defined $\hat{\bm{b}}=\mathbf{U}^{\dagger}\hat{\bm{a}}$.
Therefore Eq. (\ref{eq:ResUnity_NCGO}) in polar coordinates is:
\begin{eqnarray*}
\int\hat{\Lambda}_{N}\left(\mathbf{h}\right){\cal N}(\mathbf{h})d\mathbf{h} & = & C\int d\mathbf{U}\Delta^{2}(\mathbf{\mathbf{\bm{\lambda}}})\prod_{j=1}^{M}\left[\frac{e^{-\frac{1}{2}{\rm \lambda_{j}}}:\exp\left[\hat{b}_{j}^{\dagger}\left(e^{\bm{\lambda}}-1\right)_{j}\hat{b}_{j}\right]:}{2\cosh\left(\lambda_{j}/2\right)^{p+1}}\right]d\lambda_{j}.
\end{eqnarray*}
We will now focus on the integral over the radial part. Using the
expression of the operator term defined in Eq. (\ref{eq:Operator_term}),
we obtain:
\begin{eqnarray*}
\int\hat{\Lambda}_{N}\left(\mathbf{h}\right){\cal N}(\mathbf{h})d\mathbf{h} & = & C\int d\mathbf{U}\Delta^{2}\left(\mathbf{\bm{\lambda}}\right)\prod_{j=1}^{M}\left(1+\hat{b}_{j}^{\dagger}\left(e^{\lambda_{j}}-1\right)\hat{b}_{j}\right)\left[\frac{e^{-\lambda_{j}/2}}{2\cosh\left(\lambda_{j}/2\right)^{p+1}}\right]d\lambda_{j}\\
 & = & C\int d\mathbf{U}\Delta^{2}\left(\mathbf{\bm{\lambda}}\right)\prod_{j=1}^{M}\left(\frac{e^{-\lambda_{j}/2}}{\cosh\left(\lambda_{j}/2\right)^{p+1}}+\frac{\hat{b}_{j}^{\dagger}\hat{b}_{j}\sinh\left(\lambda_{j}/2\right)}{\cosh\left(\lambda_{j}/2\right)^{p+1}}\right)d\lambda_{j}.
\end{eqnarray*}

Here we notice that the Vandermonde determinant term is not an even
function of each eigenvalue. Hence, we do not obtain a resolution
of unity following simple parity arguments with an even weight function,
as in the previous section. The reason is that when expanding the
terms of the Vandermonde determinant, $\prod_{i<j}\left(\lambda_{i}-\lambda_{j}\right),$
for different values of $i$ and $j$, we eventually obtain operator
terms $\hat{b}_{j}^{\dagger}\hat{b}_{j}$ that depend on even \emph{and}
odd functions of $\lambda_{j}$ for a fixed value of $p$. 

On the other hand, if ${\cal N}(\mathbf{h})$ is not unitarily invariant,
and includes terms that when multiplied by the Vandermonde determinant
give an even function of $\lambda_{i}$, then we reach a different
conclusion. An example of this is if the weight function has the form:
\begin{equation}
{\cal N}(\mathbf{h})=\left[\prod_{1\le i<j\leq M}\left(\lambda_{i}+\lambda_{j}\right)^{2}\right]e^{-p\sum_{i}\lambda_{i}^{2}}
\end{equation}
In this case, it is clear that:
\begin{equation}
\Delta^{2}\left(\mathbf{\bm{\lambda}}\right){\cal N}(\mathbf{h})=\left[\prod_{1\le i<j\leq M}\left(\lambda_{i}^{2}-\lambda_{j}^{2}\right)^{2}\right]e^{-p\sum_{i}\lambda_{i}^{2}}\,.
\end{equation}

With such a weight, the integral corresponding to the operator terms
in Eq. (\ref{eq:ResUnity_NCGO}) vanishes, because of the parity of
the functions: the function $\sinh\left(\lambda_{j}/2\right)$ is
odd while the other are even functions of $\lambda_{j}$. In this
case we can obtain a resolution of unity just as in the nonstandard
symmetry case. Therefore, for the number-conserving Gaussian operators,
we can obtain a resolution of unity if the normalization factor cancels
the parity violation of the Vandermonde terms. Thus, we have shown
that an expansion of the identity operator is also possible with the
number-conserving Gaussians. As this is not a simple trace or determinant,
we cannot easily express this in terms of the original coordinates
$\mathbf{h}$. These arguments obviously do not exclude other routes
to obtaining a resolution of unity with these operators.

\section{Summary\label{sec:Summary_Conclusion}}

We have studied a subset of the general fermionic Gaussian operators,
the hermitian positive definite fermionic Gaussian operators, which
belong to a non-standard symmetry class of random matrices. We have
proved that there are simple resolutions of unity for these operators,
for any of the non-standard symmetry classes, and for any integrable
even distribution of eigenvalues. This resolution of unity is defined
for the entire Hilbert space, not for a coset or subset space as in
the case of the fermionic coherent states. 

Our proof is based on considering the hermitian Gaussian operators
as an operator basis depending on a continuous hermitian matrix. We
use polar coordinates for skew-symmetric matrices, which lead to an
expression for the resolution of unity in terms of integrals over
eigenvalues. Our result appears similar to related expressions in
random matrix theory. In order to obtain this unitarily-invariant
expression for the resolution of the unity, the normalizing factor
must be an even function of the eigenvalues. This suggests that distributions
of random matrices in the nonstandard symmetry classes, if used to
average over canonical density matrices, should generally have other
types of distribution. Otherwise the physical behaviour will simply
correspond to a unit density matrix. While this is not impossible,
it is unlikely to be generic. In the case of the symmetry class D,
we give the values of the constants corresponding to the angular and
radial integrals, with two different options for the normalizing factor.

These resolutions of unity for the Gaussian operators can be used
to derive mathematical identities for other physical applications.
As an example, our results can be used to construct a positive fermionic
distribution or fermionic Q-function. This will be carried out elsewhere.
\begin{acknowledgments}
We acknowledge useful discussions with P. J. Forrester, A. Altland
and M. Zirnbauer. L. E. C. R. Z. acknowledges financial support from
CONACYT, Mexico. P. D. D. thanks the Australian Research Council for
funding via a Discovery grant.
\end{acknowledgments}
\appendix

\section{integral over the angular coordinates\label{sec:Angular_Integral}}

In this Appendix we evaluate the integral over the angular coordinates
$\mathbf{U}$, whose expression is given in Eq. (\ref{eq:Angular_Integral}),
for the non-standard symmetry class D. 

We wish to evaluate the constant $C^{\mathbf{U}}=\int\left(\mathbf{U}^{\dagger}d\mathbf{U}\right)$.
This is the integral over the angular variables $\mathbf{U}$. It
is the angular part of the Jacobian of the matrix transformation from
Cartesian coordinates $\mathbf{H}=i\mathbf{X}$ to polar coordinates
$\left(\mathbf{U},\bm{\lambda}\right)$:
\begin{eqnarray}
\int d\mathbf{H}P\left(\mathbf{H}^{2}\right) & = & \int\left(\mathbf{U}^{\dagger}d\mathbf{U}\right)\int d\mathbf{\bm{\lambda}}\Delta^{2}\left(\bm{\lambda}^{2}\right)\prod_{j=1}^{M}P\left(\lambda_{j}^{2}\right).\label{eq:Res_Unity_PC}
\end{eqnarray}
 The value of the integral over the angular variables $\mathbf{U}$
is given by the ratio of an integral over cartesian coordinates $\mathbf{H}$,
to an integral over radial coordinates $\bm{\lambda}$:
\begin{eqnarray}
C^{\mathbf{U}} & = & \frac{\int d\mathbf{H}P\left(\mathbf{H}^{2}\right)}{\int d\mathbf{\bm{\lambda}}\Delta^{2}\left(\bm{\lambda}^{2}\right)P\left(\lambda_{j}^{2}\right)}.\label{eq:Constant}
\end{eqnarray}
 Since the constant $C^{\mathbf{U}}$ is evaluated as a ratio, we
will perform the calculations using the Gaussian form of the normalization
function. This gives simple integrals in Cartesian matrix coordinates.
For these calculations, we therefore use:
\begin{equation}
P\left(\mathbf{H}^{2}\right)=\exp\left[-p{\rm Tr}\left[\mathbf{H}^{2}\right]\right].
\end{equation}
Consequently, we now have to evaluate:
\begin{eqnarray}
C^{\mathbf{U}} & = & \frac{\int d\mathbf{H}\exp\left[-p{\rm Tr}\left[\mathbf{H}^{2}\right]\right]}{\int d\mathbf{\bm{\lambda}}\Delta^{2}\left(\bm{\lambda}^{2}\right)e^{-2p\sum_{j}\lambda_{j}^{2}}}.\label{eq:Constant_Exp}
\end{eqnarray}

\subsection*{Radial integrals}

The radial integral over the eigenvalues $\lambda_{j}$ in Eq. (\ref{eq:Constant_Exp})
is a Selberg type integral related to the Laguerre polynomials \cite{Mehta_RM_book}:
\begin{eqnarray}
\int_{-\infty}^{\infty}\ldots\int_{-\infty}^{\infty}\left|\Delta\left(x^{2}\right)\right|^{2\gamma}\prod_{j=1}^{n}\left|x_{j}\right|^{2\tilde{\alpha}-1}\exp\left(-x_{j}^{2}/2\right)d\mathbf{x} & = & 2^{\tilde{\alpha}n+\gamma n\left(n-1\right)}\prod_{j=1}^{n}\frac{\Gamma\left(1+j\gamma\right)\Gamma\left(\tilde{\alpha}+\gamma\left(j-1\right)\right)}{\Gamma\left(1+\gamma\right)}.
\end{eqnarray}
In our case, we consider $\gamma=1$, $\tilde{\alpha}=\frac{1}{2}$,
so that the radial integration gives:
\begin{eqnarray}
\int_{-\infty}^{\infty}d\mathbf{\bm{\lambda}}\Delta^{2}\left(\bm{\lambda}^{2}\right)e^{-2p\sum_{j}\lambda_{j}^{2}} & = & \left(2p\right)^{-M\left(M-1\right)}\prod_{j=1}^{M}\Gamma\left(1+j\right)\Gamma\left(j-\frac{1}{2}\right).\label{eq:Int_Radial_var}
\end{eqnarray}

\subsection*{Cartesian integrals}

The next step is evaluate the integral over the $2M\times2M$ hermitian
matrix $\mathbf{H}$, defined as:
\begin{equation}
\mathbf{H}=\left(\begin{array}{cc}
h & \Delta\\
-\Delta^{*} & -h^{T}
\end{array}\right),
\end{equation}
 with $h_{\alpha\beta}=h_{\beta\alpha}^{*}$ and $\Delta_{\alpha\beta}=-\Delta_{\beta\alpha}$.
We consider the integral over the matrices $h$ and $\Delta$ so that:
\begin{eqnarray}
\int d\mathbf{H}\exp\left[-Tr\left(p\mathbf{H}^{2}\right)\right] & = & \int\prod_{i=1}^{M}dh_{ii}\prod_{i<j}dh_{ij}^{x}dh_{ij}^{y}d\Delta_{ij}^{x}d\Delta_{ij}^{y}\exp\left[-p{\rm Tr}\left(\mathbf{H}^{2}\right)\right]\,.
\end{eqnarray}
The trace of $\mathbf{H}^{2}$ can be written as:
\begin{eqnarray}
{\rm Tr}\left[\mathbf{H}^{2}\right] & = & \sum_{ij}\left|H_{ij}\right|^{2}\nonumber \\
 & = & 2\sum_{i}^{M}h_{ii}^{2}+4\sum_{i<j}^{M}\left[\left(h_{ij}^{x}\right)^{2}+\left(h_{ij}^{y}\right)^{2}+\left(\Delta_{ij}^{x}\right)^{2}+\left(\Delta_{ij}^{y}\right)^{2}\right]\,.
\end{eqnarray}

Hence, we can use the result that:
\begin{eqnarray}
\int_{-\infty}^{\infty}dhe^{-2ph^{2}} & = & \sqrt{\frac{\pi}{2p}},\label{eq:Gaussian_Int}
\end{eqnarray}
to obtain the overall Cartesian integral of:
\begin{eqnarray}
\int d\mathbf{H}\exp\left[-pTr\left(\mathbf{H}^{2}\right)\right] & = & \left(\sqrt{\frac{\pi}{2p}}\right)^{M\left(2M-1\right)}2^{-M\left(M-1\right)},\label{eq:Int_HermitianMatrix}
\end{eqnarray}
where we have integrated over both the diagonal and the off-diagonal
terms.

\subsection*{Normalization constant}

Using the results of Eq. (\ref{eq:Int_Radial_var}) and Eq. (\ref{eq:Int_HermitianMatrix})
we obtain:
\begin{eqnarray}
C^{\mathbf{U}} & = & \pi^{M\left(M-\frac{1}{2}\right)}2^{-M\left(M-1\right)}\prod_{j=0}^{M-1}\frac{1}{\Gamma\left(2+j\right)\Gamma\left(j+\frac{1}{2}\right)}\,.\label{eq:Constant1}
\end{eqnarray}

\end{document}